\documentclass[runningheads]{llncs}
\usepackage[T1]{fontenc}
\usepackage{graphicx}
\usepackage{booktabs}
\usepackage{subcaption}
\usepackage{mwe}
\usepackage{wrapfig}
\usepackage{makecell}
\usepackage{float}
\usepackage{algorithm2e}
\usepackage{siunitx}
\usepackage[accsupp]{axessibility}
\usepackage[misc,geometry]{ifsym}
\usepackage{tikz}
\usepackage{hyperref}
\usepackage{color}

\usepackage{ifthen}
 \newcommand{\method}{\textsc{PyramidAI}\xspace}
\newcommand{\positiveRate}{positive retention rate}
\newcommand{\ie}{\textit{i}.\textit{e}., }

\newcommand\copyrighttext{%
    \footnotesize \textcopyright 2026 The Authors, under exclusive license to Springer Nature Switzerland AG.
    This preprint has not undergone peer review or any post-submission improvements or corrections. The Version of Record of this contribution is published in Lecture Notes in Computer Science (LNCS, volume 15902) included in the ``European Conference on Parallel Processing'' conference series, and is available online at \href{https://doi.org/10.1007/978-3-031-99872-0_21}{https://doi.org/10.1007/978-3-031-99872-0\_21}.
}
\newcommand\copyrightnotice{%
    \begin{tikzpicture}[remember picture,overlay]
        \node[anchor=south,yshift=30pt,fill=yellow!20] at (current page.south) {\fbox{\parbox{\dimexpr\textwidth-\fboxsep-\fboxrule\relax}{\copyrighttext}}};
    \end{tikzpicture}%
}

\begin{document}
\title{Efficient Pyramidal Analysis of Gigapixel Images on a Decentralized Modest Computer Cluster}
\titlerunning{Efficient Pyramidal Analysis of Gigapixel Images}
\author{Marie Reinbigler\inst{1}\textsuperscript{(\Letter)}  \and Rishi Sharma\inst{2} \and Rafael Pires\inst{2} \and Elisabeth Brunet\inst{1} \and Anne-Marie Kermarrec\inst{2} \and Catalin Fetita\inst{3}}

\institute{SAMOVAR, Inria Saclay, Télécom SudParis, IP Paris, 91120 Palaiseau, France \email{marie.reinbigler@telecom-sudparis.eu}
\and
EPFL, Lausanne, Switzerland
\and SAMOVAR, Télécom SudParis, IP Paris, 91120 Palaiseau, France}

\authorrunning{M. Reinbigler et al.}

\maketitle              %
\copyrightnotice
\begin{abstract}
  Analyzing gigapixel images is recognized as computationally demanding. In this paper, we introduce PyramidAI, a technique for analyzing gigapixel images with reduced computational cost.
  The proposed approach adopts a gradual analysis of the image, beginning with lower resolutions and progressively concentrating on regions of interest for detailed examination at higher resolutions. %
   We investigated two strategies for tuning the accuracy-computation performance trade-off when implementing the adaptive resolution selection, validated against the Camelyon 16 dataset of biomedical images. Our results demonstrate that PyramidAI substantially decreases the amount of processed data required for analysis by up to $2.65\times$, while preserving the accuracy in identifying relevant sections on a single computer. To ensure democratization of gigapixel image analysis, we evaluated the potential to use mainstream computers to perform the computation by exploiting the parallelism potential of the approach. Using a simulator, we estimated the best data distribution and load balancing algorithm according to the number of workers. The selected algorithms were implemented and highlighted the same conclusions in a real-world setting. %
   Analysis time is reduced from more than an hour to a few minutes using 12 modest workers, offering a practical solution for efficient large-scale image analysis.

\end{abstract}

\section{Introduction}
\label{sec:intro}

Gigapixel images are omnipresent in various fields like biomedicine or satellite imagery. Featuring a pyramidal multi-resolution structure, they reach sizes of up to $10^5 \times 10^5$ pixels at the highest resolution, equivalent to \SI{40}{\giga\byte} of uncompressed data~\cite{Xu:2017}. Among their benefits, they hold immense potential for advancing biomedical research~\cite{Israeli:2019,Childers:2014,Gurcan:2009} by providing unprecedented detail and clarity.  
The counterpart is a significant computational cost, raising challenges for traditional image analysis methods, especially in processing large datasets efficiently.
Deep learning techniques, including convolutional neural networks (CNNs)~\cite{LeCun:1989} and vision transformers~\cite{Kolesnikov:2021}, have shown remarkable results in image analysis~\cite{Atabansi:2023}. 
These methods ensure a thorough examination that minimizes the chance of missing essential details as it analyzes the entire image. While effective, they are resource-intensive and not always computable on local facilities.

Typical gigapixel image analysis is initiated with a preliminary coarse detection which identifies the analysis area, and then divides it into tiles for processing. These tiles can be analyzed either at their highest resolution alone ~\cite{Kassani:2019,Iizuka:2020,Amin-Naji:2019,Komura:2018} or in conjunction with their contextual lower-resolution counterparts \cite{Muhammad:2018,Wetteland:2020,Abdeltawab:2021,Rijthoven:2021,Schmitz:2021,Xiang:2022}.
Our proposal aims at further enhancing computational efficiency, leveraging the inherent multiresolution structure of these images. 
By filtering tiles requiring %
in-depth analysis, PyramidAI preserves computational resources and, thanks to its potential for load distribution, enables optimization of computational load while maintaining analytical rigor. Thus, our contributions are:
\begin{itemize}
	\item the design of \method, a pyramidal analysis approach tailored for gigapixel images. 
 This iterative process ensures that only areas of interest are examined at the highest resolution;
    \item  two strategies to determine the threshold of the binary decision to proceed with analysis to a higher resolution level for a specific tile;
    \item an experimental evaluation using the Camelyon16 dataset~\cite{Jama:2017}, comprising biomedical images. The analysis focuses on detecting metastatic cells. 
    The results demonstrate that \method significantly optimizes efficiency by reducing the average number of tiles analyzed by up to $2.65\times$ while successfully identifying $90\%$ of the true positive tiles detected through the conventional highest resolution analysis across the entire test set. 
    \item a proposition of load balancing in a distributed setting was simulated for a frugal and democratized access to gigapixel image analysis.
\end{itemize}

After contextualization in §\ref{related_works}, \method{} design and components are detailed in §\ref{pyramidai}, its application to the Camelyon 16 dataset and the evaluation compared to the highest resolution only analysis are in §\ref{evaluation}. We explore its  potential for computation distribution in §\ref{Eval_distrib} before concluding with some perspectives in~§\ref{conclusion}.

\section{Related work}
\label{related_works}

Due to input size constraints, large multiresolution images are traditionally divided into tiles that can be fed into models for analysis. This approach is commonly utilized for the analysis of biomedical images~\cite{Kassani:2019,Iizuka:2020,Amin-Naji:2019}, as well as in the field of satellite imaging~\cite{Li:2018,deJong:2019}. To better integrate contextual information, previous works have investigated the use of lower resolutions by either analyzing multiple resolutions and aggregating the outcomes or selecting an optimal compromise resolution~\cite{Muhammad:2018,Wetteland:2020,Abdeltawab:2021,Schmitz:2021,Xiang:2022} with the goal of maximizing the accuracy. Yet those approaches become prohibitively computationally expensive with the increasing size of images. 

In this paper, we take an orthogonal approach and focus on computation efficiency. With the advent of Transformer neural network architecture~\cite{Vaswani:2017} and its adaptation to vision tasks~\cite{Kolesnikov:2021}, new categories of analysis emerged in both domains~\cite{Atabansi:2023,Adegun:2023}, which strengthen the trend toward increasingly more computation-intensive analysis~\cite{Khan:2022}. Thus, limiting the computation to areas of interest is more important than ever. For example, the Viola-Jones detector~\cite{viola:2001} processes tiles at one resolution, but applies a cascade of classifiers only on retained tiles from previous tests. Similarly, background removal detects tiles candidates prone to information, such as tissues~\cite{Babbar:2019,Huang:2023}, by leveraging one low-resolution level, which sketches a coarse pyramidal approach. The effectiveness of exploiting several lower-resolution levels has been explored in the work of~\cite{Goffe:2010}, which introduced an image segmentation algorithm that begins at a low resolution and selectively zooms in to refine the segmentation boundaries. Here, we expanded on the pyramidal approach, applying it to a broader range of tasks, including AI-driven analysis, and proposed generic methods for zoom-in decisions.

The heavy compute load of an exhaustive execution is handled by parallelizing known tile analysis load among distributed workers with a central controller \cite{ParallelExec:2012}. Since the pyramidal execution tree size is not known in advance, a dynamic balancing strategy is needed to minimize execution time. Work stealing is commonly used for its dynamic adaptability to different configurations and load characteristics \cite{WorkStealingFoundation:1999,AsyncWorkStealing:2024,WorkStealingDynamic:2021}.

\section{Pyramidal analysis design}
\label{pyramidai}

\begin{figure}[ht]
	\centering
	\includegraphics[width=0.9\textwidth]{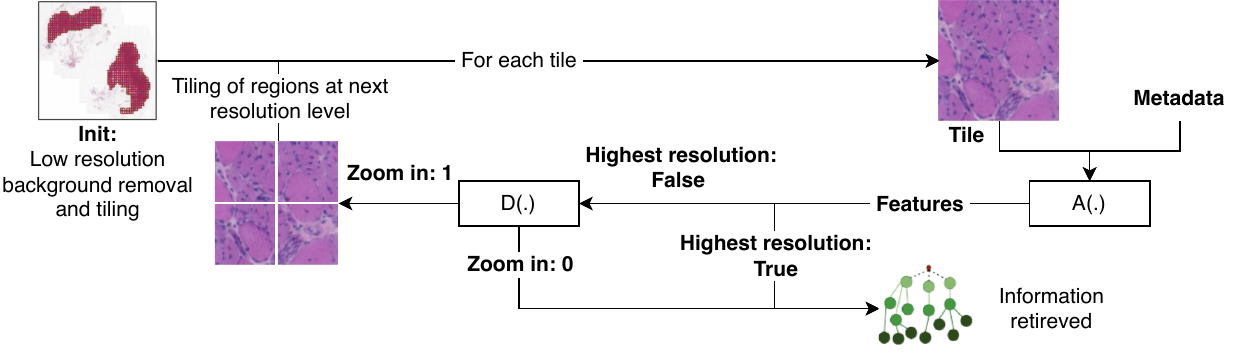}
	\caption{The analysis starts at a low resolution, for each tile an analysis block $A(.)$ extracts features. Based on them, a decision block $D(.)$ decides to either stop the analysis at that level for that tile or to zoom-in and process the corresponding tiles at a higher resolution. Extracted features can be further exploited for downstream tasks such as whole slide prediction or segmentation.}
	\label{fig:method_workflow}
\end{figure}

Pyramidal analysis aims to reduce the total amount of processed data by leveraging the different resolution levels of gigapixel images. By selectively analyzing portions of the image, the algorithm trades accuracy for computation performance. 
This section first presents the pyramidal analysis algorithm (§\ref{algo}), illustrated in Figure~\ref{fig:method_workflow}, and describes two strategies to select per-level zoom-in decision thresholds (§\ref{thresh_meth}), each offering distinct accuracy-performance trade-off. 

\subsection{The pyramidal analysis algorithm}
\label{algo}

Once the input tile size is determined, the analysis begins at a low, pre-selected image resolution. Tiles of interest are progressively filtered from one resolution level to the next, up to the highest resolution. Indeed, at each resolution level, an analysis block denoted by $A(.)$ is applied to each remaining tile, potentially supplemented with metadata information tailored to the specific use case, \ie{} tile position in the image.
The output features are fed into a decision block $D(.)$, whose binary outcome determines whether the pyramidal analysis continues at the next resolution level.
Tiles identified as areas of interest are subdivided into higher resolution and start a new iteration of the algorithm. The \textit{scale factor} $f$ of the input multiresolution image indicates that a tile at level $R_n$ corresponds to $f^2$ new tiles of the same dimension at level $R_{n-1}$, with $R_0$  being the highest and $R_N$ the lowest resolution.
The analysis and decision blocks are tailored to the specific use case and adapted for each resolution to optimize accuracy.

The efficiency of the pyramidal analysis depends on the density of relevant information in the image at low resolution. Lower densities imply greater speedup as more tiles will be discarded. If most of the tiles are of interest, intermediate pyramid levels filter only a few tiles, \ie{} most of the intermediate resolution tiles as well as the highest resolution ones are analyzed. This will lead to poorer performance than the reference analysis at the highest resolution only.
In practice, this issue is limited as the maximum slowdown $S$ is bounded by Equation (\ref{slowdown}) for a pyramid with an infinite number of levels, and remains relatively small. Our evaluation (§\ref{evaluation}) demonstrates that the speedup is greater than 1 on the Camelyon 16 dataset for a wide range of decision thresholds.

\vspace{-1.3em}
\begin{equation}
	\label{slowdown}
	S(f) = \sum^{\infty}_{R_n = 0}\frac{1}{f^{2R_n}} = \frac{f^2}{f^2 - 1} \hspace{2cm}
	S(2) = \frac{4}{3} \approx 1.33 \,; \hspace{0.3cm}
	S(3) = \frac{9}{8} = 1.125
\end{equation}
Given the size of gigapixel images, the number of tiles to analyze can be considerable, especially in the case of a single-worker execution.
The tile-based approach has great potential for load distribution as the analysis can be performed in parallel independently from the tile location or resolution level. However, by design, the total computational load is unknown in advance and increases exponentially when a zoom-in is performed. Thus, a dynamic load balancing policy is needed to distribute the workload among workers at runtime, which is studied in §\ref{Eval_distrib}.

\subsection{Decision block threshold selection strategies}
\label{thresh_meth}

The decision block $D(.)$ determines if a zoom-in is required for a particular tile based on the analysis output. $D(.)$ acts as a classifier whose decision threshold needs to be tuned to achieve the desired accuracy-performance trade-off. As for the analysis block, $D(.)$ requires per-resolution parameters, \ie{} one threshold per resolution level. The proper tuning of each threshold is crucial to the overall performance and accuracy of \method{}, as false negatives can miss areas of interest while false positives can degrade performance.

To ease the tuning of each decision threshold, we propose two strategies for threshold selection. The first maximizes computation performance given an objective user-defined metric expressed as a retention rate of a reference value; here, it denotes the proportion of true areas of interest retained with our pyramidal approach compared to the ones detected by the highest resolution only analysis. The second allows the user to choose a suitable accuracy-performance trade-off through a single graph.
Both strategies rely on the $F_\beta$ score (eq.~\ref{fscore}), the weighted harmonic mean between precision and recall: a higher $\beta$ favors recall over precision, improving the true positives at the expense of false positives. To tune the thresholds, we first compute the $F_\beta$ score for all resolution levels on the train set. This requires collecting the predictions for all tiles of all resolution levels. Then, for a given $\beta$ we select the threshold for resolution level $R_n$ such that it maximizes $F_\beta$, \textit{i.e.} $argmax_{t \in [0, 1]}(F_\beta(t))$. This can be approximated by maximizing $F_\beta$ over a finite set of sampled thresholds. %

\textbf{Metric-based threshold selection.} That strategy aims to maximize the speedup under the constraint of an objective minimal retaining rate for a user-defined metric. Starting from a user-provided minimal retention rate $r$, the strategy consists of maximizing the speedup on a per-resolution level basis such that the isolated impact on the retention rate for each level (among $n$) is at least the $n$-th root of $r$. The isolated impact is computed by considering a pyramidal execution where all resolution levels are pass-through except the level of interest, and results in decision tables similar to Figure \ref{fig:LevelRetentionnRate} in §\ref{metric_based_thresh}. 
Using this strategy, $\beta$ values are selected independently for each resolution level as the smallest $\beta$, ensuring an objective retention rate.
This ensures that the global retention rate, which in the worst case is the product of the $n$-th individual retention rates, is at least $r$. 

\textbf{Empirical threshold selection.} 
That strategy offers greater flexibility, allowing the user to empirically determine the trade-off between performance and the retaining of the user-defined metric. 
Based on retrieved predictions, for each $\beta$ value the pyramidal execution is computed on each gigapixel image of the train set using the thresholds corresponding to the same $\beta$ at all resolution levels. Based on those executions, the reduction in the number of tiles analyzed, indicative of the speedup, and the retention rate for the final metric compared to the analysis at the highest resolution only  can be estimated for each $\beta$ value. Based on those figures, an empirical selection of $\beta$ can be made based on a single graph, similar to Figure \ref{fig:EmpiricalFig}~(\emph{a}) presented in §\ref{evaluation}.

\vspace{-1em}
\begin{equation}
	\label{fscore}
	F_\beta = (1+\beta^2)\frac{Precision*Recall}{(\beta^2*Precision)+Recall} = \frac{(1+\beta^2)*TP}{(1+\beta^2)*TP+ \beta^2*FN+FP}
\end{equation}

\section{\method experiments}
\label{evaluation}

We evaluate \method{} on the Camelyon 16 challenge dataset~\cite{Jama:2017}, using the metric-based (§\ref{metric_based_thresh}) or the empirical (§\ref{fexible_thresh}) thresholds, by comparing against the reference scenario, which analyzes all tiles at the highest resolution after background detection. %
Section~\ref{Dataset} introduces Camelyon 16 use-case, followed by the description of analysis blocks at each resolution level in §\ref{models}, 
the computation time measurement strategy in §\ref{computationTime} and the impact of these thresholds on the whole image classification accuracy in §\ref{wholeSlideClassification}.

\subsection{Camelyon use-case and data preprocessing}
\label{Dataset}

The Camelyon~\cite{Jama:2017} dataset is composed of Hematoxylin–Eosin(HE)-stained sentinel lymph node histological images coming from breast cancer patients' biopsies. The training set contains 160 healthy slides and 110 with nodal metastases. 129 additional whole-slide images (49 with and 80 without metastases) are available for test purposes. Those images can reach sizes of about $10^5\times2.10^5$ pixels, \ie{} up to \SI{80}{\giga\byte} of uncompressed data, and include up to 10 resolution levels.

In our experiments, tiles of 224x224 pixels are extracted after a background removal using Otsu thresholding technique \cite{Otsu:1979} performed using \cite{Lu:2021clam} tool. Stain normalization was applied using Macenko method \cite{Macenko:2009} and \cite{stainlib:2022} tool.

The analysis block A(.) detects the probability of the presence of metastasis in the tile and uses deep learning image classification models detailed in §\ref{models}.
The decision block D(.) is a threshold function with a threshold to be determined according to §\ref{thresh_meth} strategies.
The final metric to preserve is the ratio of true positive tiles retained at the highest resolution by our pyramidal approach versus the ones detected by the reference execution, further denoted by \textit{\positiveRate}.
Our pyramidal analysis is based on a 3-level pyramid, the highest resolution being level 0 and the lowest being level 2, with a scale factor of 2. 

\subsection{Analysis blocks models}
\label{models}

\begin{figure}[ht]
  \centering
  \begin{subfigure}[t]{0.23\textwidth}
    \includegraphics[width=\textwidth]{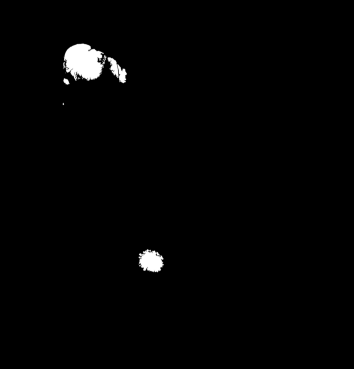}
    \caption{Ground truth.}
  \end{subfigure}
  \hfill
  \begin{subfigure}[t]{0.23\textwidth}
    \includegraphics[width=\textwidth]{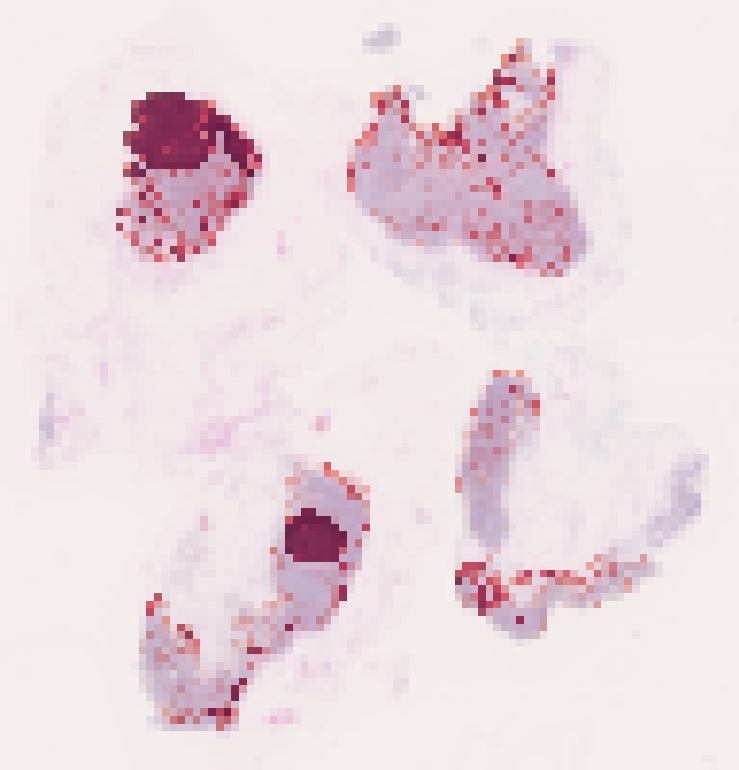}
    \caption{Lowest pyramid resolution.}
  \end{subfigure}
  \hfill
  \begin{subfigure}[t]{0.23\textwidth}
    \includegraphics[width=\textwidth]{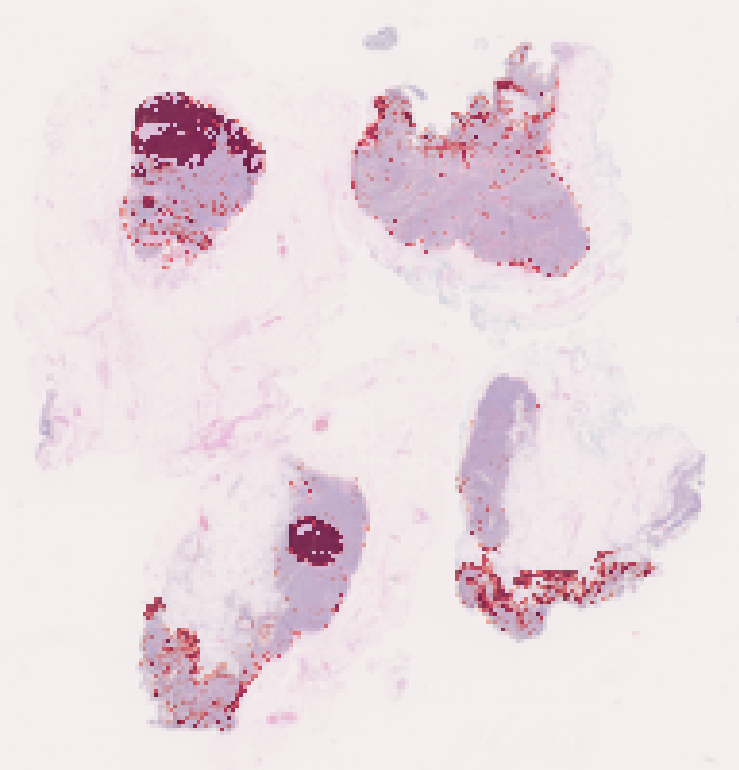}
    \caption{Intermediate pyramid resolution.}
  \end{subfigure}
  \hfill
    \begin{subfigure}[t]{0.23\textwidth}
    \includegraphics[width=\textwidth]{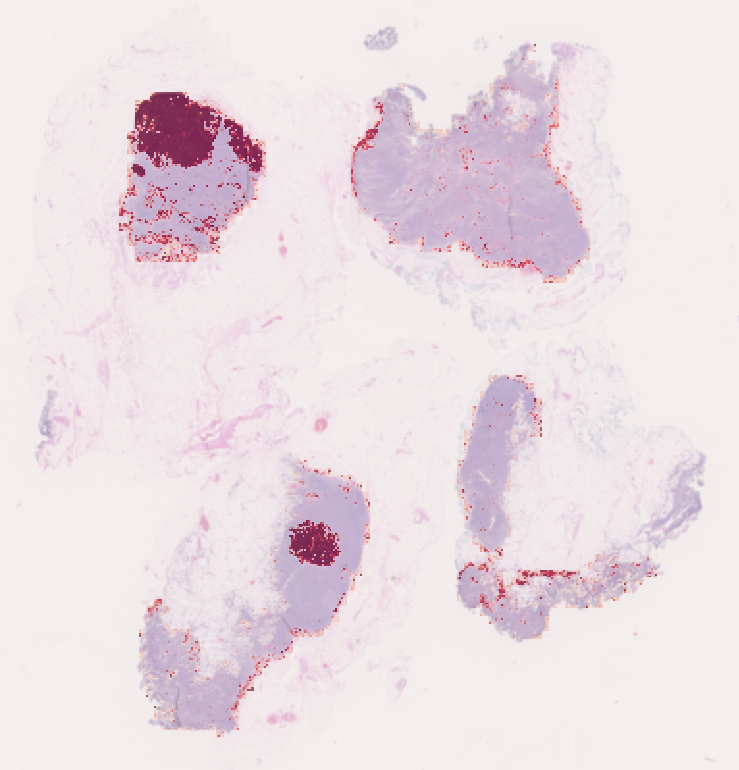}
    \caption{Highest pyramid resolution.}
  \end{subfigure}
      \setlength\belowcaptionskip{-0.4cm}

  \caption{Model tumor probability heatmap compared to ground truth. The darker the red, the higher the probability.}
  \label{fig:ModelsPerf}
\end{figure}

A model was trained for each resolution level in a supervised manner. Each dataset contains up to 192000 tiles extracted from  Camelyon 16 whole-slide images, split with 80/20 for training and validation, and preprocessed (see §\ref{Dataset}). The training set is balanced by keeping all tumoral tiles and randomly selecting the same number of normal tiles. The test set is a random subset of pre-processed tiles extracted from Camelyon 16 test slides. Dataset sizes are summarized in Table~\ref{tab:DatasetSizes}. Online data augmentation (random flips and rotations) was used. The model architecture is based on the InceptionV3 model \cite{Szegedy:2016} with a GlobalAverage2D pooling layer, a dense layer with a depth of 224, and a final sigmoid layer. The weights were randomly initialized for resolutions 0 and 1. Resolution 2 uses initialized Imagenet weights and the model was obtained via a transfer learning process. 
Training was based on accuracy, on the  Adam optimizer with a learning rate of 1e-4. 100 epochs were set up with 1000 steps per epoch. The final model accuracies are summarized in Table~\ref{tab:ModelsAcc}. The model performance can be qualitatively assessed in Figure~\ref{fig:ModelsPerf}.

\begin{table}[ht]
	\centering
	\begin{minipage}[t]{0.45\textwidth}
		\begin{tabular}{|c|c|c|c|}
            \hline
			&	\thead{Train \\set size}	&	\thead{Validation \\set size}	&	\thead{Test \\set size}	\\
			\hline
			Level 0 	&	26576	&	38400	&	92000	\\
            \hline
			Level 1	&	26134	&	38400	&	92000	\\
            \hline
			Level 2	&	25504	&	38400	&	72568	\\
			\hline
		\end{tabular}

		\caption{Train, validation, and \\test set sizes for each resolution.}
		\label{tab:DatasetSizes}
	\end{minipage}
	\begin{minipage}[t]{0.45\textwidth}
		\begin{tabular}{|c|c|c|c|}
			\hline
			&	\thead{Train \\accuracy}	&	\thead{Validation \\accuracy}	&	\thead{Test \\accuracy}	\\
			\hline
			Level 0 	&	0.9328	&	0.9498	&	0.9480	\\
            \hline
			Level 1	&	0.9439	&	0.9590	&	0.9584	\\
            \hline
			Level 2	&	0.8982	&	0.9110	&	0.9166	\\
			\hline
			
		\end{tabular}

		\caption{Train, validation, and test accuracies for each resolution model}
  
		\label{tab:ModelsAcc}
	\end{minipage}
 
	\label{ModelsTrained}  
\end{table}

\subsection{Computation time measurement}
\label{computationTime}

Image pre-processing, dataset creation, and inferences were executed on the IDRIS supercomputer Jean-Zay. Model training ran on a server with 2 AMD EPYC 7502 with 32 cores, 512 GB of RAM, and a Nvidia Quadro RTX 5000 GPU. Time measurements were performed on a mainstream computer embedding an Intel Core i5-9500 with 6 cores and 16GB of RAM.

Due to the large computation time in an isolation environment induced by the large number of tiles in the test set, which will be analyzed by our method and the reference, we estimated computation time based on the following measurements. The pyramidal analysis has three phases. The first one is the initialization, the retrieval of all tiles at the lowest resolution after background removal. Then, tiles are analyzed by the resolution-specific analysis block, and the decision block is applied. Finally, new tasks are created and added to the working queue if a zoom-in decision is positive. For the reference analysis, only initialization time and analysis block computation time at the highest resolution are considered. 
For the initialization phase, we measured computation time 1000 times per slide, and for the other phases 1000 times on one slide as it is tile-dependent. Results are gathered in Table~\ref{tab:ComputationTime}. 

Based on the data and the models' prediction probabilities on the test set tiles, we can simulate "post-mortem" computation for reference and pyramidal analysis knowing the total number of tiles per resolution level that have been analyzed. Task creation time and initialization are not considered in the simulation as the analysis blocks computation time is dominant. 

\begin{table}[ht]
	{%
	\centering
	\begin{tabular}{|c|c|c|c|c|c|}
        \hline
    	& initialization & \thead{Level 0 \\analysis block} & \thead{Level 1 \\analysis block} & \thead{Level 2 \\analysis block} & Task creation\\
    	\hline
    	Time (s) & 0.02$\pm$0.01 & 0.33$\pm$0.04 & 0.33$\pm$0.03 & 0.31$\pm$0.03 & 2.77$\pm$0.89e-5 \\
        \hline
	\end{tabular}
	}
  \setlength\belowcaptionskip{-0.5cm}

     \caption{Computation time per phase}%
 \label{tab:ComputationTime}
\end{table}

\subsection{Decision blocks tuned with metric-based thresholds}
\label{metric_based_thresh}

We followed the first methodology from §\ref{thresh_meth} for applications where a minimum value for a metric needs to be guaranteed. 

\begin{figure}[htbp]
	\includegraphics[width = \textwidth]{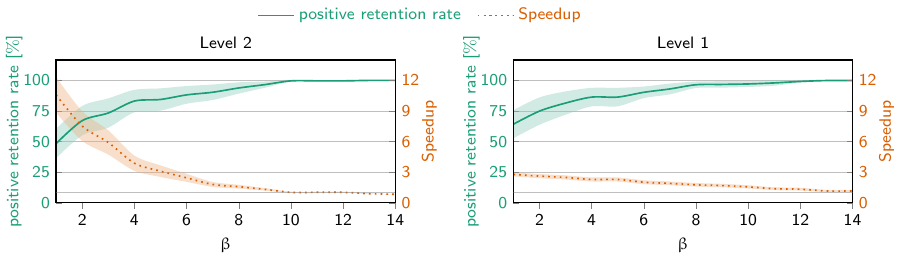}
  \setlength\belowcaptionskip{-0.5cm}
\vspace{-0.6cm}
	\caption{Influence of each isolated resolution level on \positiveRate{} and speedup.}
    \label{fig:LevelRetentionnRate}

\end{figure}

Thirty slides from the training set are tiled at each resolution level, starting with the lowest resolution. Then, the corresponding trained model is applied to each tile of each resolution level and the inference result, i.e. the probability for the tile to contain a tumor, is collected.
These probabilities are used to measure precision and recall values for several thresholds. For $\beta$ values ranging from 1 to 14, we select the threshold maximizing the $F_\beta$ score. Then, on each slide, we isolate each resolution level, meaning we apply the threshold corresponding to each $\beta$ score for this resolution and zoom in everywhere in all other resolution levels, to measure its impact on the \positiveRate{} at the highest resolution. We average this \positiveRate{} between all thirty slides. Results are summarized in Figure~\ref{fig:LevelRetentionnRate}.

Given an objective \positiveRate{} for our application, as we have two intermediate resolution levels, we take the square root of the objective value, which will be the objective \positiveRate{} to be reached for each isolated resolution level. We select the lowest $\beta$ value that achieves the objective \positiveRate{} at the highest resolution. 

As noticed in Figure~\ref{fig:MetricFig}, the technique enables to reach the objective value expected on the test set too. For instance, given an objective \positiveRate{} of 0.90, it results in an intermediate level \positiveRate{} to be reached of 0.9487, which is achieved at resolution 1 for $\beta$=8 and by $\beta$=9 at resolution 2 (Fig.~\ref{fig:LevelRetentionnRate}). 
92\% of true positive tiles of the reference analysis are retained with our approach
with 2.34 times fewer tiles analyzed, \ie{} a speedup of 2.34. These values highlight also the benefit of using multiple-level pyramidal analysis. Indeed, the global objective of \positiveRate{} of 0.90 considering isolated resolution 1 is reached for $\beta$=6 with only 2.02 times fewer tiles analyzed; considering resolution 2 only, the objective is reached with $\beta$=7 with 1.80 times fewer tiles analyzed. Combining resolution levels helps to reduce tiles analyzed while ensuring the same \positiveRate{}. 
Using a global objective equal to 0.90, following §\ref{computationTime} methodology, we estimated average computation time per slide on the test set to about 1h19min $\pm$1h09min, compared to 2h29min $\pm$1h34min for reference execution. The huge standard deviation is explained by the significant variation in the number of tiles per slide by up to a factor of 30.

\subsection{Decision blocks tuned with empirical thresholds}
\label{fexible_thresh}

As in §\ref{metric_based_thresh}, we measure the precision and recall values for several thresholds, and, for $\beta$ values ranging from 1 to 14, we select the threshold maximizing the $F_\beta$ score. The pyramidal analysis is performed using thresholds corresponding to the same $\beta$ value for each resolution level in the decision blocks. For each $\beta$, the \positiveRate{} and the speedup were measured.

\begin{figure}[t]
\noindent\begin{minipage}[t]{0.33\textwidth}
  \centering
    \includegraphics[height=3.7cm]{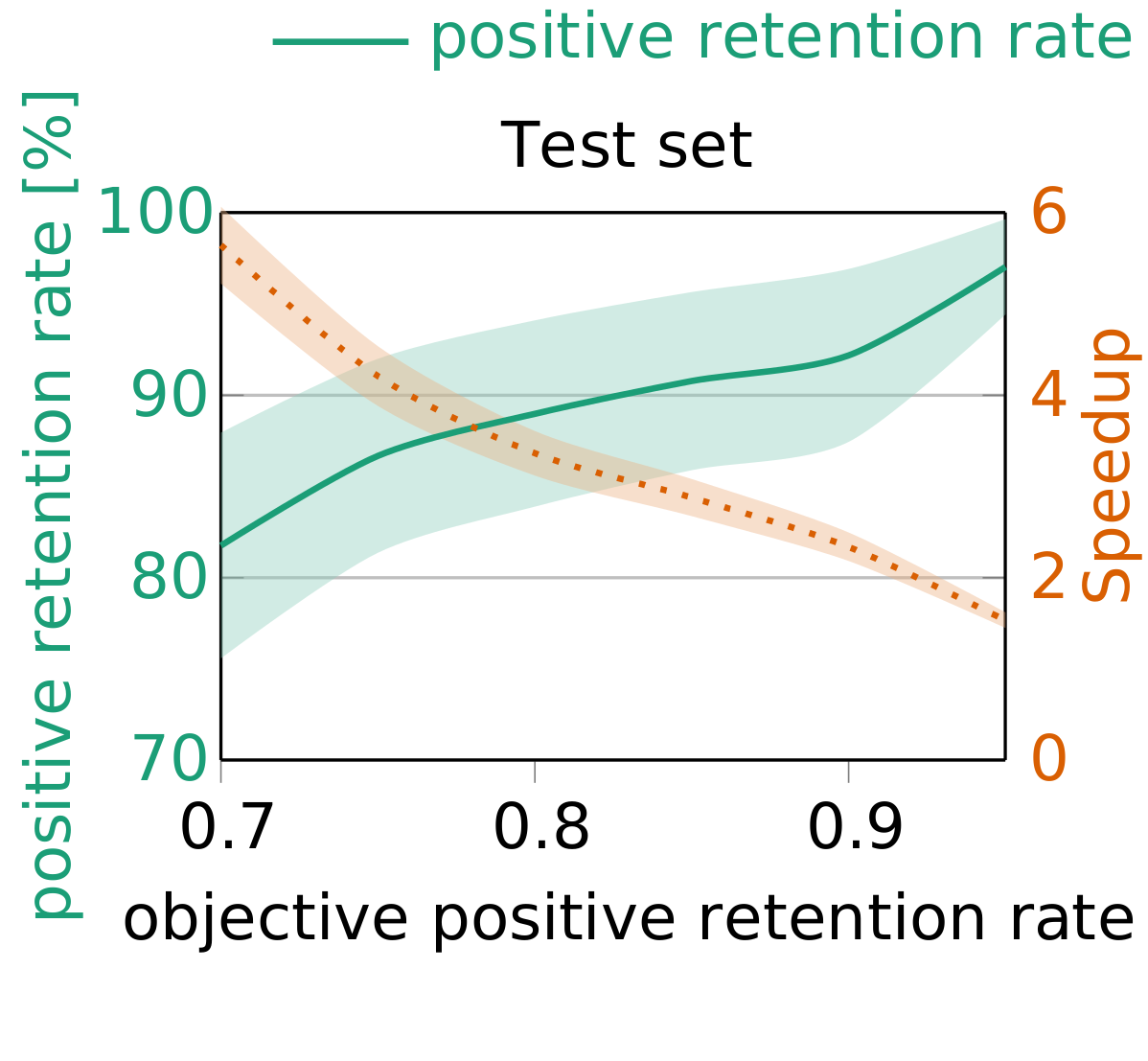}
    \vspace{-0.4cm}
    \caption{Trade-off according to an objective positive retention rate (cf. §\ref{metric_based_thresh})}
    \label{fig:MetricFig}
\end{minipage}
\hspace{0.01\textwidth}
\noindent\begin{minipage}[t]{0.64\textwidth}
  \centering
    \includegraphics[height=3.7cm]{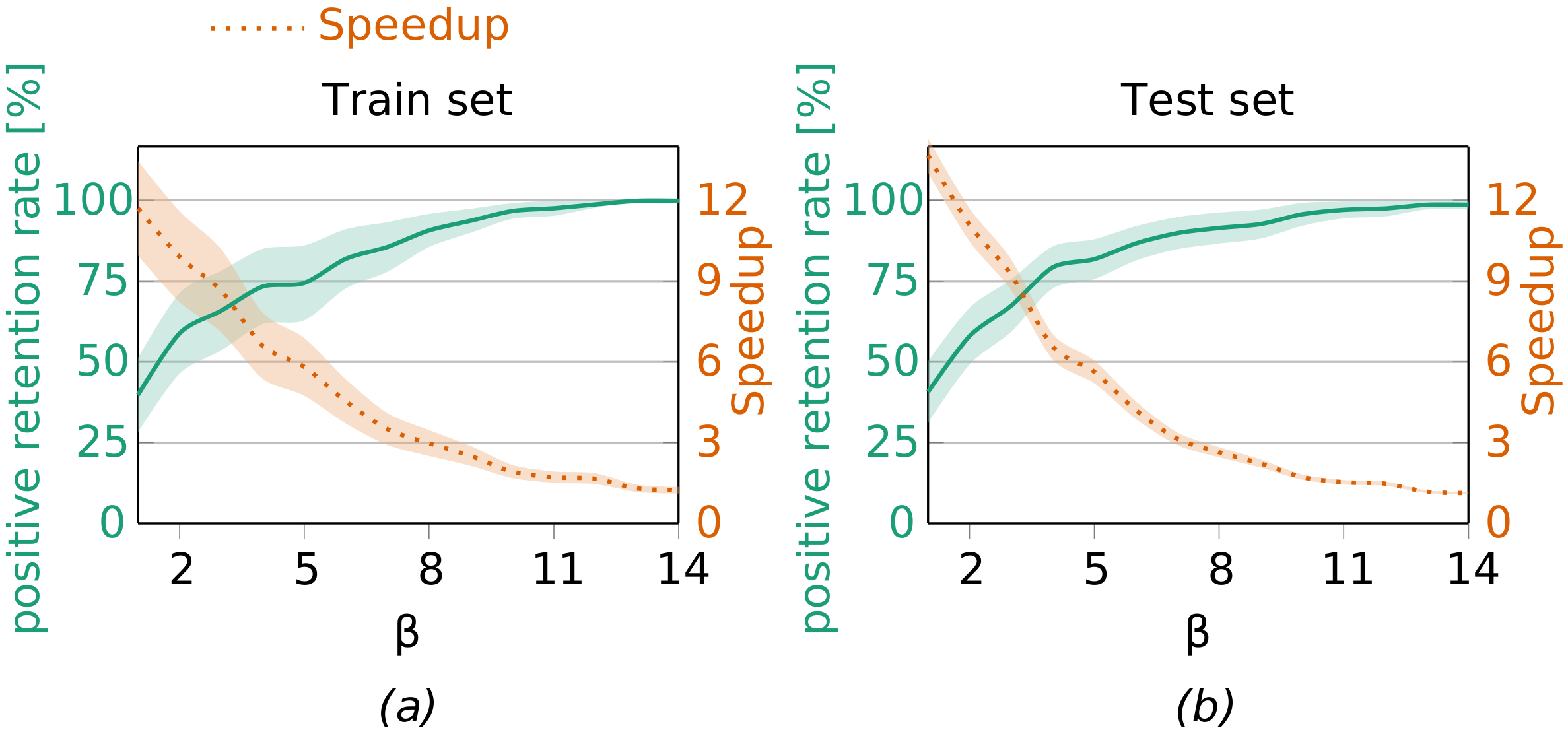}
    \vspace{-0.4cm}
    \caption{Trade-off between \positiveRate{} and speedup in a pyramidal execution according to empirical thresholds of $\beta$ (cf. §\ref{fexible_thresh})}
    \label{fig:EmpiricalFig}
\end{minipage}
\vspace{-0.3cm}
\setlength\belowcaptionskip{-0.5cm}

\end{figure}

Figures~\ref{fig:EmpiricalFig}~(\emph{a}) and (\emph{b}) summarize the results for our use-case, allowing the user to easily decide which trade-off between accuracy-performance to choose by selecting only one $\beta$ value. This is very useful in applications where losing some accuracy is acceptable or can be compensated by a post-processing step. Figure~\ref{fig:EmpiricalFig} outlines the potential of the pyramidal approach to drastically speed up the analysis. For example, given a $\beta$ value of 5, retaining 80\% of positive tiles on the train set, more than 80\% of positive tiles of the test set are retained with 5.63$\times$ fewer tiles analyzed. For applications sensitive to accuracy reduction as ours, we select the $\beta$ value retaining 90\% of positive tiles on the train set, naming $\beta$=8, which retains 90\% of the test set positive tiles, with a speedup of 2.65.
Following §\ref{computationTime} methodology, the average computation time per slide on the whole test set is estimated to be about 1h11min $\pm$1h06min.

\subsection{Whole slide image classification}
\label{wholeSlideClassification}

We assessed the impact of positive tile loss using PyramidAI on the whole-slide image (WSI) classification.
We trained a bagging decision tree classifier to predict tumoral images from the distribution of tile prediction probabilities.
When stopping predictions at a lower resolution level with \method{}, we projected the predicted probability onto all corresponding tiles at the highest resolution.
The baseline, without \method{}, achieves an accuracy of 0.84.
We obtain the same accuracy of 0.84 with a 2.65x speedup when using \method{} with an empirical threshold.
The metric-based strategy achieves a lower accuracy of 0.77, due to the strategy optimized for true positives retention, and detects tumor presence on 35 WSI slides versus 29 for the baseline, but at the cost of a higher false positive rate.

\section{Distributed \method experiments}
\label{Eval_distrib}

We assess the data distribution and load balancing policies adapted to our pyramidal approach execution for a decentralized cluster of mainstream computers. Section~\ref{SimuDescr} presents the implementation of a simulator to estimate the most promising combination of a data distribution strategy and a load-balancing policy. Section~\ref{SimuWithSynchro} explores synchronization-based load-balancing policy while §\ref{SimuWithoutSynchro} assesses the potential of removing synchronization and using a work-stealing-based balancing policy. Section~\ref{RealExecution} describes the implementation of the chosen policy deployed on a real cluster of mainstream computers.

\subsection{Simulation implementation}
\label{SimuDescr}

As stated in §\ref{computationTime}, most of the time is spent in the analysis blocks execution. Thus, to estimate the load per worker, we consider the maximum number of tiles computed per worker to evaluate the load balancing. Based on the pyramidal execution tree retrieved using thresholds from §\ref{fexible_thresh}, we simulate an execution offline based on the number of workers, the initial tile distribution on a low-resolution level, and the load balancing policy.
The considered initial tile distributions on the lowest resolution are: the \textbf{Round-Robin distribution} consisting in iterating over low-resolution tiles and dispatching cyclically one tile per worker until it is exhausted; the \textbf{Random distribution} where we shuffle low-resolution tile list and dispatch data by block of balanced size to each worker; and the \textbf{Block distribution} with a sorted low-resolution tile list by location in the image which is dispatched by block of balanced size among workers.

We compared data distribution combined with load balancing policies to the ideal scenario where the oracle knows in advance which tiles and how many will be analyzed at each resolution level and thus dispatches them in a balanced way independently of the resolution level. This is the lowest execution time achievable. We also compare to the previous references: the execution time at highest resolution only, and the pyramidal approach on a single worker. The results correspond to an average maximum number of tiles analyzed per worker for the entire test set.

\subsection{Load balancing via synchronizations}
\label{SimuWithSynchro}

\begin{figure}[ht]
  \centering
  \begin{subfigure}[t]{0.49\textwidth}
    \includegraphics[width=\textwidth]{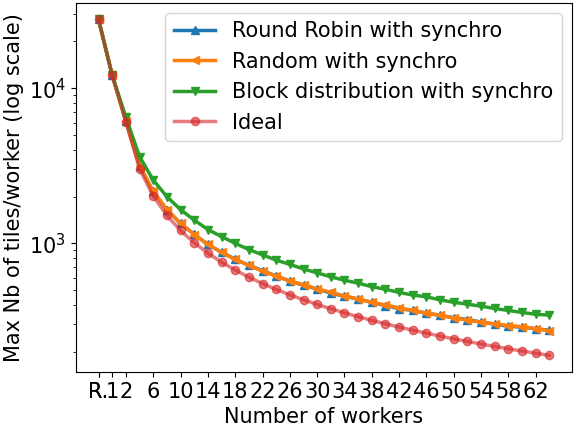}
    \caption{With synchronisation.}
    \label{fig:withSynchro}
  \end{subfigure}
  \hfill
    \begin{subfigure}[t]{0.49\textwidth}
    \includegraphics[width=\textwidth]{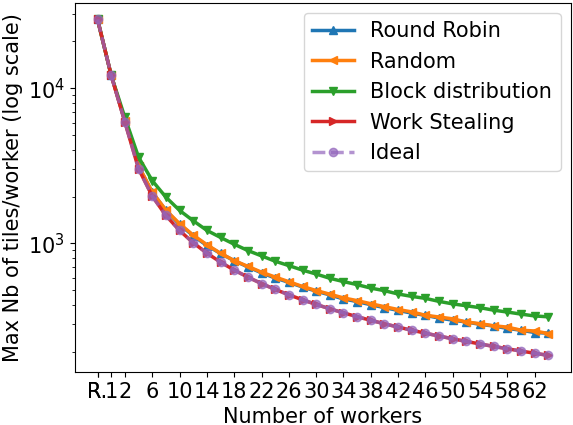}
    \caption{Without synchronisation.}
    \label{fig:withoutSynchro}
  \end{subfigure}
  \setlength\belowcaptionskip{-1cm}
  \caption{Maximum number of tiles analyzed by the busiest worker depending on data distribution strategies and load balancing policies (work stealing or synchronization). R. refers to the number of tiles analyzed for the reference execution on one worker, \ie{} the highest resolution only analysis.}
  \label{ModelsPerf}
\end{figure}

As a first naive approach, we balance the load after each resolution level before jumping to the next  analysis level, meaning after all low-resolution tiles are analyzed and after all intermediate-resolution tiles pass. We combine this policy with each data distribution strategy. Figure~\ref{fig:withSynchro} presents the results. From this graph, we can deduce that Round-Robin and Random data distribution strategies provide similar results, with a higher stability for the Round-Robin one. The distribution by blocks of tiles localized in the same region appears to be inefficient, which is explained by the heterogeneity of the distribution of tumoral tissues in the image. 

\subsection{Dynamic scheduling approach without synchronization}
\label{SimuWithoutSynchro}

Because the tiles analysis is independent of each other and of the resolution level, we explored the potential of synchronization-free approaches. 
We tested all data distribution strategies without load rebalancing at runtime, 
and the Round-Robin approach combined with the dynamic work-stealing policy. 
The work-stealing strategy consists of randomly choosing a neighbor to be the victim of the work-stealing once the working queue is empty. If it has more than one task remaining, it dequeues one task corresponding to a leaf of the current pyramidal execution graph state and transmits it to the sender. Otherwise, it returns an empty list of tasks and the thief chooses another victim. In the simulation, message transfer time is neglected as  compute time is dominant over message transfer time. 

Figure~\ref{fig:withoutSynchro} confirms that the Round-Robin technique is the most stable one and that even without dynamic load balancing policies, the maximum load of the busiest worker is close to the one with synchronization and should be favored for its simplicity. With an increasing number of workers, the considered work-stealing method is the most efficient, especially starting with 4 workers, and is equivalent to the ideal case as message passing latency is neglected.

\subsection{Deployment on a cluster of modest computers}
\label{RealExecution}

\begin{figure}[ht]
\centering
    \includegraphics[width=0.55\textwidth]{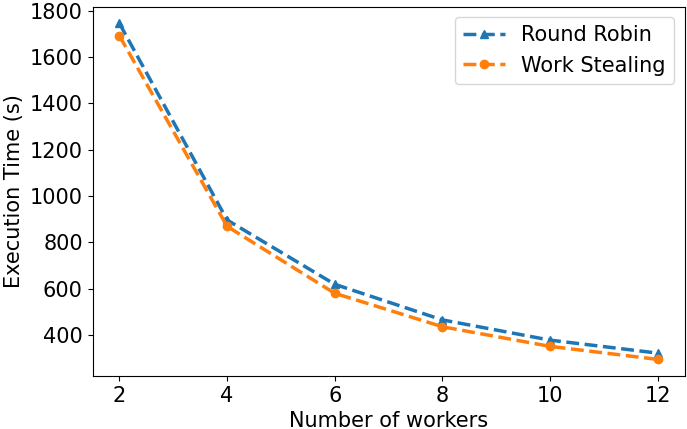}
     \setlength\belowcaptionskip{-1cm}

  \caption{Average execution time per image evaluated on real data using a Round-Robin data distribution according to the number of workers, with and without work-stealing.}
  \label{fig:RealExecPerf}
\end{figure}

Simulation results demonstrated that Round-Robin is the most efficient low-resolution tile distribution strategy. For the dynamic load balancing policy, work-stealing is increasingly beneficial with the number of workers, even if close to Round-Robin only with less than 4 workers. We validate these conclusions with a cluster of 12 fully-connected mainstream computers equipped with an Intel Core i5-9500 and 16GB of RAM. Data is replicated among workers without shared memory, allowing compatibility with any topology and extension to subimage replication. The implementation uses the DecentralizePy~\cite{DecentralizePy:2023} framework for TCP connections among workers, compatible with decentralized settings and any topology. Each worker has its own queue and uses work-stealing previously described when it is empty. Indeed, it requests a task by message-passing to a neighbor. If the victim has remaining tasks, it sends one back. Otherwise, it sends an empty message. The victim updates its list of potential victims as the sender worker runs out of tasks already. Finally, all workers send their subtrees, including stolen subtrees, back to node 0 for full tree reconstruction and further processing. The conclusions are verified on three images: one with large tumors, one with several small ones, and one negative image, each measure being computed 3 times. Figure \ref{fig:RealExecPerf} confirms that using the work-stealing technique is adequate for this workload, especially with a growing number of workers. 

\section{Conclusion \& Perspectives }
\label{conclusion}

In this paper, we propose a pyramidal approach for gigapixel image analysis, reducing computation compared to the highest resolution-only analysis, suitable for constrained environments. It leverages the multiresolution feature of gigapixel images by starting analysis at a low resolution, tiling tissue regions, and applying an analysis block to each tile. The extracted features feed a decision block  that classifies tiles as of interest and proceeds the analysis at the next higher resolution level when required. The decision criterion is a key element to ensure a good trade-off between the global accuracy of the pyramidal analysis and its computational performance compared to the reference. Thus, we propose two strategies to tune decision block thresholds per resolution level. Our methods, evaluated on the Camelyon 16 dataset, demonstrated we could divide by up to 2.65 the number of tiles to be analyzed while detecting 90\% of positive tiles among those detected by reference execution. Computation time decreased from over 2 hours to 1 hour per image on average.
\method{}'s potential to adapt to a modest computing environment is highlighted, especially when combined with a dynamic work-stealing policy. It provides results for an image in about 15 minutes instead of in more than an hour using 12 mainstream computers. It democratizes access to such analysis for scientists with limited computational resources.
Though illustrated on a gigapixel biomedical use case, the approach is generalizable to any gigapixel images, such as satellite or spatial images.

\begin{credits}
\subsubsection{\ackname} The authors acknowledge the financial support of \href{https://www.hi-paris.fr}{Hi! PARIS Center}. and the use of HPC resources from GENCI–IDRIS (Grant 2023-AD010614264)

\subsubsection{\discintname}
The authors have no competing interests to declare that are relevant to the content of this article.
\end{credits} 

\bibliographystyle{splncs04}
\bibliography{egbib-short}

\begin{thebibliography}{10}
\providecommand{\url}[1]{\texttt{#1}}
\providecommand{\urlprefix}{URL }
\providecommand{\doi}[1]{https://doi.org/#1}

\bibitem{Abdeltawab:2021}
Abdeltawab, H., et~al.: A pyramidal deep learning pipeline for kidney
  whole-slide histology images classification. Scientific Reports  (2021)

\bibitem{Adegun:2023}
Adegun, A., Viriri, S., Tapamo, J.R.: Review of deep learning methods for
  remote sensing satellite images classification: experimental survey and
  comparative analysis. Journal of Big Data  (2023)

\bibitem{Amin-Naji:2019}
Amin-Naji, M., Aghagolzadeh, A., Ezoji, M.: Ensemble of cnn for multi-focus
  image fusion. Information Fusion  (2019)

\bibitem{Atabansi:2023}
Atabansi, C., et~al.: A survey of transformer applications for
  histopathological image analysis: New developments and future directions.
  BioMedical Engineering OnLine  (2023)

\bibitem{Babbar:2019}
Babbar, J., Rathee, N.: Satellite image analysis: A review. In: IEEE
  International Conference on Electrical, Computer and Communication
  Technologies (2019)

\bibitem{Jama:2017}
Bejnordi, B., et~al.: {Diagnostic Assessment of Deep Learning Algorithms for
  Detection of Lymph Node Metastases in Women With Breast Cancer}. JAMA  (2017)

\bibitem{WorkStealingFoundation:1999}
Blumofe, R.D., Leiserson, C.E.: Scheduling multithreaded computations by work
  stealing. J. ACM  (1999)

\bibitem{ParallelExec:2012}
Bueno, G., et~al.: A parallel solution for high resolution histological image
  analysis. Computer Methods and Programs in Biomedicine  (2012)

\bibitem{Childers:2014}
Childers, M., et~al.: Gene therapy prolongs survival and restores function in
  murine and canine models of myotubular myopathy. Science translational
  medicine  (2014)

\bibitem{DecentralizePy:2023}
Dhasade, A., et~al.: Decentralized learning made easy with decentralizepy. In:
  EuroMLSys'23. ACM (2023)

\bibitem{AsyncWorkStealing:2024}
Fernandes, J.B., et~al.: Adaptive asynchronous work-stealing for distributed
  load-balancing in heterogeneous systems. arXiv:2401.04494  (2024)

\bibitem{WorkStealingDynamic:2021}
Freitas, V., et~al.: Packsteallb: A scalable distributed load balancer based on
  work stealing and workload discretization. Journal of Parallel and
  Distributed Computing  (2021)

\bibitem{Goffe:2010}
Goffe, R., Damiand, G., Brun, L.: A causal extraction scheme in top-down
  pyramids for large images segmentation. In: SSPR (2010)

\bibitem{Gurcan:2009}
Gurcan, M.N., et~al.: Histopathological image analysis: A review. IEEE Reviews
  in Biomedical Engineering  (2009)

\bibitem{Huang:2023}
Huang, P.W., et~al.: Deep-learning based breast cancer detection for
  cross-staining histopathology images. Heliyon  (2023)

\bibitem{Iizuka:2020}
Iizuka, O., et~al.: Deep learning models for histopathological classification
  of gastric and colonic epithelial tumours. Scientific Reports  (2020)

\bibitem{Israeli:2019}
Israeli, D., et~al.: An {AAV}-{SGCG} {Dose}-{Response} {Study} in a
  $\gamma$-{Sarcoglycanopathy} {Mouse} {Model} in the {Context} of {Mechanical}
  {Stress}. Molecular Therapy. Methods \& Clinical Development  (2019)

\bibitem{deJong:2019}
de~Jong, K.L., Sergeevna~Bosman, A.: Unsupervised change detection in satellite
  images using convolutional neural networks. In: IJCNN (2019)

\bibitem{Kassani:2019}
Kassani, S., et~al.: Classification of histopathological biopsy images using
  ensemble of deep learning networks. In: Computer Science and Software
  Engineering (2019)

\bibitem{Khan:2022}
Khan, S., et~al.: Transformers in vision: A survey. ACM Comput. Surv.  (2022)

\bibitem{Kolesnikov:2021}
Kolesnikov, A., et~al.: An image is worth 16x16 words: Transformers for image
  recognition at scale (2021)

\bibitem{Komura:2018}
Komura, D., Ishikawa, S.: Machine learning methods for histopathological image
  analysis. Computational and Structural Biotechnology Journal  (2018)

\bibitem{LeCun:1989}
LeCun, Y., et~al.: Backpropagation applied to handwritten zip code recognition.
  Neural Computation  (1989)

\bibitem{Li:2018}
Li, Y., et~al.: Deep learning for remote sensing image classification: A
  survey. WIREs Data Mining and Knowledge Discovery  (2018)

\bibitem{Lu:2021clam}
Lu, M.Y., et~al.: Data-efficient and weakly supervised computational pathology
  on whole-slide images. Nature Biomedical Engineering  (2021)

\bibitem{Macenko:2009}
Macenko, M., et~al.: A method for normalizing histology slides for quantitative
  analysis. In: IEEE Symposium on Biomedical Imaging: From Nano to Macro (2009)

\bibitem{Muhammad:2018}
Muhammad, N., et~al.: A {Multi}-resolution {Deep} {Learning} {Framework} for
  {Lung} {Adenocarcinoma} {Growth} {Pattern} {Classification} (2018)

\bibitem{stainlib:2022}
Ot{\'a}lora, S., et~al.: stainlib: a python library for augmentation and
  normalization of histopathology h\&e images. bioRxiv  (2022)

\bibitem{Otsu:1979}
Otsu, N.: A threshold selection method from gray-level histograms. IEEE
  Transactions on Systems, Man, and Cybernetics  (1979)

\bibitem{Rijthoven:2021}
Rijthoven, M., et~al.: {HookNet}:{Multi}-resolution convolutional neural
  networks for semantic segmentation in histopathology whole-slide images.
  Medical Image Analysis  (2021)

\bibitem{Schmitz:2021}
Schmitz, R., et~al.: Multi-scale fully convolutional neural networks for
  histopathology image segmentation: From nuclear aberrations to the global
  tissue architecture. Medical Image Analysis  (2021)

\bibitem{Szegedy:2016}
Szegedy, C., et~al.: Rethinking the inception architecture for computer vision.
  In: IEEE conference on computer vision and pattern recognition (2016)

\bibitem{Vaswani:2017}
Vaswani, A., et~al.: Attention is all you need. In: Advances in Neural
  Information Processing Systems (2017)

\bibitem{viola:2001}
Viola, P., Jones, M.: Rapid object detection using a boosted cascade of simple
  features. In: IEEE conference on computer vision and pattern recognition
  (2001)

\bibitem{Wetteland:2020}
Wetteland, R., et~al.: A {Multiscale} {Approach} for {Whole}-{Slide} {Image}
  {Segmentation} of five {Tissue} {Classes} in {Urothelial} {Carcinoma}
  {Slides}. Technology in Cancer Research \& Treatment  (2020)

\bibitem{Xiang:2022}
Xiang, T., et~al.: Dsnet: A dual-stream framework for weakly-supervised
  gigapixel pathology image analysis. IEEE Transactions on Medical Imaging
  (2022)

\bibitem{Xu:2017}
Xu, Y., et~al.: Large scale tissue histopathology image classification,
  segmentation, and visualization via deep convolutional activation features.
  BMC Bioinformatics  (2017)

\end{thebibliography}

\end{document}